\documentstyle[prl,aps,epsfig,twocolumn]{revtex}

\begin{document}

\wideabs{

\title{Enhancement of Kerr nonlinearity via multi-photon coherence}

\author{
    A.~B.~Matsko$^{1}$,
    I.~Novikova$^{1}$,
    G.~R.~Welch$^{1}$,
    and M.~S.~Zubairy$^{1,2}$
}
\address{
    $^{1}$Department of Physics and Institute for Quantum Studies,
    Texas A\&M University, College Station, Texas 77843-4242
}
\address{
    $^{2}$Department of Electronics, Quaid-i-Azam University,
    Islamabad, Pakistan
}


\date{\today}
\maketitle

\begin{abstract}

    We propose a new method of resonant enhancement
    of optical Kerr nonlinearity using multi-level
    atomic coherence.  The enhancement is accompanied
    by suppression of the other linear and nonlinear
    susceptibility terms of the medium.  We show that
    the effect results in a modification of the nonlinear
    Faraday rotation of light propagating in an $^{87}$Rb
    vapor cell by changing the ellipticity of the light.

\end{abstract}


}
\tighten
\noindent

    Efficient manipulation with optical quanta calls
for materials possessing large and lossless optical
nonlinearities~\cite{qq1}.  Recently it was shown that coherent
atomic effects such as electromagnetically induced transparency
(EIT)~\cite{EIT} and coherent population trapping
(CPT)~\cite{CPT,sz} are able to suppress the linear absorption of
resonant multilevel media while keeping the nonlinear
susceptibility at a very high
level~\cite{harris90prl,largenl,luknature}. Previous experimental
and theoretical work has shown that coherent media produce an
effective interaction between two electromagnetic fields due to
both refractive~\cite{largenl,luknature,harris99prl,lukin00prl}
and absorptive~\cite{absorption,absorption_exp} $\chi^{(3)}$ Kerr
nonlinearities.

    One method of producing Kerr nonlinearity with vanishing
absorption is based on the coherent properties of a three-level
$\Lambda$ configuration (see Fig.~\ref{fig1}a), in which the
effect of EIT can be observed.  However, since an ideal CPT medium
does not interact with the light, it also cannot produce any
nonlinear effects~\cite{sz}.  To get a nonlinear interaction in
such a medium one needs to ``disturb'' the CPT regime by
introducing, for example, the interaction with an additional
off-resonant level (the $N$-type scheme, Fig.~\ref{fig1}b). If the
disturbance of CPT is small, i.e., the detuning $\Delta$ is large,
the absorption does not increase much, but the nonlinearity can be
as strong as in a near-resonant two level
system~\cite{largenl,luknature,lukin00prl,zubairy01pra}.

    In this Letter we propose a new method of resonant
enhancement of $\chi^{(3)}$ and higher order nonlinearities
without significant optical losses.  We consider the $M$-type
configuration shown in Fig.~\ref{fig1}c.  There is coherent
population trapping in such a scheme which leads to EIT. At the
same time, by introducing a small two-photon detuning $\delta$ we
may disturb this CPT and produce a strong nonlinear coupling among
the electromagnetic fields interacting with the atomic system.

    An important advantage of the $M$-type level
configuration is that this type of coherence is easily created on
the Zeeman sublevels of alkali atoms.  In this particular case
large self-phase modulation of circularly polarized light is
possible.  We show here significant change in nonlinear
magneto-optical polarization rotation~\cite{faraday} due to
$M$-type coherence in Rb vapor.  Our experimental results confirm
the theoretical predictions.

    Our method of creation of a highly nonlinear medium
with small absorption has prospects in fundamental as well as
applied physics.  An advantage of the $M$ configuration is that by
increasing the number of the levels it is possible to realize
higher orders of nonlinearity.  This can be used for construction
of non-classical states of light as well as coherent processing of
quantum information.

    Let us first consider a medium (atomic, molecular,
semiconductor, etc.) with an $M$-type energy level structure as in
Fig.~\ref{fig1}c.  Here levels $|a_j\rangle$ have natural decays
$\gamma_j$ and we assume that the ground state levels
$|b_j\rangle$ have no decay.  The coherence between levels
$|b_i\rangle$ and $|b_j\rangle$ ($i \ne j$) have slow homogeneous
decay $\gamma_{0}$.  The energy levels are coupled by weak probe
electromagnetic fields of Rabi frequency $\alpha_j$ and strong
coupling fields of Rabi frequency $\Omega_j \gg |\alpha_j|$.  For
the sake of simplicity we assume all the fields to be resonant
with the corresponding atomic transitions except for the coupling
field $\Omega_2$ which has small detuning $\delta \ll \gamma_2$
from the $|b_3\rangle \rightarrow |a_2\rangle$ transition.

    We first focus on the case when there is no coherence
decay in the system. Then the interaction of the atoms and the
fields can be described in the slowly varying amplitude and phase
approximations by the Hamiltonian
\begin{eqnarray} \nonumber
H_M = -\hbar \delta |b_3 \rangle \langle b_3| + \hbar \sum
\limits_{j=1}^{2} ( \alpha_j |a_j \rangle \langle b_j| + \Omega_j
|a_j \rangle \langle b_{j+1}| + H.c.),
\end{eqnarray}
where $H.c.$ means Hermitian conjugation.  If $\delta=0$, there
exists the noninteracting eigenstate (``dark
eigenstate''~\cite{sz}) of this Hamiltonian, corresponding to the
zero eigenvalue $\lambda_{D}=0$:
\begin{eqnarray} \label{d2}
|D\rangle =
 \frac{\alpha_1\alpha_2 |b_3\rangle - \Omega_2
\alpha_1 |b_2\rangle + \Omega_1\Omega_2 |b_1 \rangle}{
\sqrt{|\alpha_1|^2 |\alpha_2|^2 + |\Omega_2|^2 |\alpha_1|^2 +
|\Omega_1|^2|\Omega_2|^2}}.
\end{eqnarray}

    Strictly speaking,  there is no ``dark state'' for
finite $\delta$.  However when the detuning is small enough
$(\delta \ll \gamma_2, |\Omega_2|^2/\gamma_2)$, the disturbance of
the ``dark'' state is small. The disturbed dark state $|\tilde D
\rangle $ and the corresponding eigenvalue of the Hamiltonian
$\lambda_{\tilde D}$ in the limit of $|\Omega_j| \gg |\alpha_j|$
are:
\begin{eqnarray} \label{eigenvect}
|\tilde D \rangle  \approx  \zeta \left [ | D \rangle - \delta
\frac {\alpha_1^*|\alpha_2|^2} {|\Omega_1|^2|\Omega_2|^2} | a_1
\rangle + \delta \frac {\alpha_1 \alpha_2}
{|\Omega_1||\Omega_2|^2} | a_2 \rangle \right ],
\end{eqnarray}
\begin{eqnarray} \label{eigenvalt}
\lambda_{\tilde D} \approx -\hbar \delta \frac
{|\alpha_1|^2|\alpha_2|^2} {|\Omega_1|^2|\Omega_2|^2},
\end{eqnarray}
where $\zeta \simeq 1$ is a normalization parameter.

    Because the system does not leave the dark state
during static or adiabatic interaction with the probe fields we
write the Hamiltonian as $H_M \simeq \lambda_{\tilde D} |\tilde D
\rangle \langle \tilde D |$. As $|\tilde D \rangle \langle \tilde
D | \simeq |b_1 \rangle \langle b_1 | \simeq 1$, it is convenient
to exclude atomic degrees of freedom from the interaction picture
and  to rewrite the Hamiltonian in Heisenberg picture with
quantized probe fields.  The relation between Rabi frequencies of
the probe fields and quantum operators describing the
corresponding field mode can be written as
\begin{equation}
\hat \alpha_i = \sqrt{\frac{2\pi \wp^2_i \nu_i}{\hbar V_i}} \hat
a_i = \xi_i \hat a_i,
\end{equation}
where $\wp_i$ is the dipole moment of the transition $|a_i\rangle
\rightarrow |b_i\rangle$, $\nu_i$ is the field frequency, $V_i$ is
the quantization volume of the mode, $\hat a_i$ and $\hat
a_i^\dag$ are the annihilation and creation operators.  Then the
Hamiltonian takes the final form
\begin{eqnarray} \label{hamM}
H_M= -\hbar \delta \frac {\xi_1^2\xi_2^2}
{|\Omega_1|^2|\Omega_2|^2} \ \hat a_1^\dag \hat a_1 \hat a_2^\dag
\hat a_2.
\end{eqnarray}
We see that the nonlinear coupling between the probe fields
increases with an increase in the detuning $\delta$.

    To understand the size of the nonlinear interaction
we recall the interaction Hamiltonian for an $N$
scheme~\cite{largenl,zubairy01pra}:
\begin{eqnarray} \label{hamN}
H_N= \hbar \frac {\xi_1^2\xi_2^2} {\Delta |\Omega_1|^2} \ \hat
a_1^\dag \hat a_1 \hat a_2^\dag \hat a_2.
\end{eqnarray}
The ratio of coupling constants in (\ref{hamM}) and (\ref{hamN})
\begin{equation}
{\cal R} = \frac{\delta \Delta }{|\Omega_2|^2}
\end{equation}
determines the relative strength of the nonlinear interaction
introduced by the schemes.  If $|{\cal R}| > 1$, then the $M$
scheme is more effective than $N$.

    However, this comparison is not consistent unless we
take into account the absorption of the probe field introduced by
spontaneous emission.  Simple calculations demonstrate that the
ratio of the spontaneous emission probabilities coincide with the
ratio of nonlinear susceptibilities for the $M$ and $N$ schemes.
Another source of probe absorption is due to the decay of
coherence between ground state levels which leads to depopulation
of the dark state created in the system and to the absorption of
the probe fields independently of each other. It can be shown that
the probability of this process is the same for both $M$ and $N$
configurations.

    In the usual cells containing gases of alkali atoms
the one-photon transitions are Doppler-broadened due to the motion
of the atoms.  If the condition for EIT is fulfilled ($\Omega_1
\gg W_d \sqrt{\gamma_0/\gamma_1}$, where $W_d$ is the Doppler
linewidth $W_d \gg \gamma_1$ \cite{leearchiv}), then in the
$M$-scheme the population of level $|b_2 \rangle$ is approximately
equal to $|\alpha_1|^2/|\Omega_1|^2$. The nonlinear interaction
results from the refraction and absorption of the second probe
field $\alpha_2$, coupled to the second drive field $\Omega_2$.
Fields $\alpha_2$ and $\Omega_2$ along with levels $|b_2\rangle$,
$|b_3\rangle$, and $|a_2\rangle$ create $\Lambda$ system.  Almost
all population is in level $|b_2 \rangle$. Therefore,
\begin{equation} \label{sus1}
\chi_M = -i\frac{3}{8\pi^2}{\cal N}\lambda_{\alpha 2}^3
\frac{\gamma_2
(\gamma_0+i\delta)}{(\gamma_0+i\delta)W_d+|\Omega_2|^2}
\frac{|\alpha_1|^2}{|\Omega_1|^2}.
\end{equation}
where ${\cal N}$ is the atomic density and $\lambda_{\alpha 2}$ is
the wavelength of the field $\alpha_2$.

    Similar calculations allow us to derive the
susceptibility for the field $\alpha_2$ for the Doppler broadened
$N$ scheme:
\begin{equation} \label{sus2}
\chi_N = -i\frac{3}{8\pi^2}{\cal N}\lambda_{\alpha 2}^3
\frac{\gamma_2}{W_d+i\Delta} \frac{|\alpha_1|^2}{|\Omega_1|^2},
\end{equation}
It is easy to see that Eqs.~(\ref{sus1}) and (\ref{sus2}) are
interchangeable if $\gamma_0 \rightarrow 0$, and $\Delta
\leftrightarrow \delta/|\Omega_2|^2$.  Therefore, ultimately $M$
and $N$ interaction schemes are equally efficient, even though
quite different mechanisms are responsible for the nonlinear
susceptibility enhancement.

    To experimentally demonstrate the enhancement of
nonlinearity in the $M$ type level scheme we study the rotation of
elliptical polarized light resonant with the $F=2 \rightarrow
F'=1$ transition of the $^{87}$Rb $D_1$ line (Fig.~\ref{fig2}a).
We consider the light as two independent circular components $E_+$
and $E_-$ which generate a coherent superposition of the Zeeman
sublevels (a dark state).  To disturb this dark state we apply a
longitudinal magnetic field $B$ which leads to a splitting $\delta
\propto B$ of the Zeeman sublevels. This atomic transition
consists of $\Lambda$ and $M$ level configurations
(Fig.~\ref{fig2}a and \ref{fig2}b).

%
%

    The nonlinear properties of the $M$ level scheme result
in significant modification of the nonlinear polarization rotation
as a function of the light ellipticity.  We write the field
amplitudes as $|E_\pm|^2 = (1\pm q)|E_0|^2/2$, where $|q|<1$
characterizes the light ellipticity and find an expression for the
rotation angle $\phi$ of the polarization ellipse for small
magnetic field.  The details of the calculation will be given
elsewhere.  The enhancement of rotation with regards to the
isolated $\Lambda$ system is given by
\begin{equation} \label{rot1}
\frac {\phi(B)}{\phi_\Lambda(B)} \approx  \frac 12 +
\frac{2+q^2}{(2-q^2)^2},
\end{equation}

    We have measured the polarization rotation in a
cell containing Rb vapor using the technique described in detail
in~\cite{novikova'01pra}.  The results are shown in
Fig.~\ref{fig3}.  The experimental dependence looks slightly
different from the theoretical one (Fig.~\ref{fig3} inset) because
of the influence of the Doppler broadening and AC-stark shifts due
to light coupling to off-resonant atomic sublevels. However,
numerical simulations based on steady state solution of exact
density matrix equations give a good agreement with the
experiment.

    The experimental results pertain to a semiclassical
description where the field is classical. An interesting
application of the Kerr nonlinearity with quantized field as given
in Eq.~(\ref{hamM}) is the possible implementation of a quantum
phase gate.  A quantum phase gate, together with a one-bit unitary
gate, form the basic building block for quantum
computation~\cite{nc}. The transformation properties of a quantum
phase gate leave the two qubit states unchanged when one or both
input qubits are in the logic state 0 and introduces a phase $\eta
$ only when both the qubits in the input states are 1.  For input
photon states $|0\rangle$ or $|1\rangle$ for the two qubits, a
unitary operator of the form $Q_{\eta}={\rm exp}(i \eta \hat
a_1^\dag \hat a_1 \hat a_2^\dag \hat a_2)$ can lead to such a
phase gate, i.e., $Q_{\eta}|0_1,0_2\rangle= |0_1,0_2\rangle$,
$Q_{\eta}|0_1,1_2\rangle= |0_1,1_2\rangle$,
$Q_{\eta}|1_1,0_2\rangle= |1_1,0_2\rangle$, and
$Q_{\eta}|1_1,1_2\rangle={\rm exp}(i \eta) |0_1,0_2\rangle$. It is
clear that such a phase gate can be realized via Hamiltonian $H_M$
with the time-evolution unitary operator $exp(-iH_Mt)$ and the
corresponding phase $\eta= \hbar \delta \xi_1^2\xi_2^2
t/|\Omega_1|^2|\Omega_2|^2$.

    The resonant enhancement of $\chi^{(5)}$ and higher
orders of nonlinearity may be achieved using additional $\Lambda$
sections connected to $M$ scheme, similar to the generalized $N$
scheme~\cite{zubairy01pra}.  Such $\chi^{(5)}$ nonlinearity may be
so high that three-photon phase gates become feasible.

    In conclusion, we have proposed a realization of media
with resonantly enhanced Kerr nonlinearity where one-photon
resonant absorption is suppressed due to coherence effects. Such
media have certain advantages over already existing schemes of
coherent resonant nonlinearity enhancement and hold promise for
the use in the creation of non-classical states of light and in
the implementation of quantum computing algorithms.

    The authors gratefully acknowledge the support from
Air Force Research Laboratory (Rome, NY), DARPA-QuIST, the Office
of Naval Research, and the TAMU Telecommunication and Informatics
Task Force (TITF) initiative.


\begin{flushleft}
\begin{figure} 
 \center{
\epsfig{file=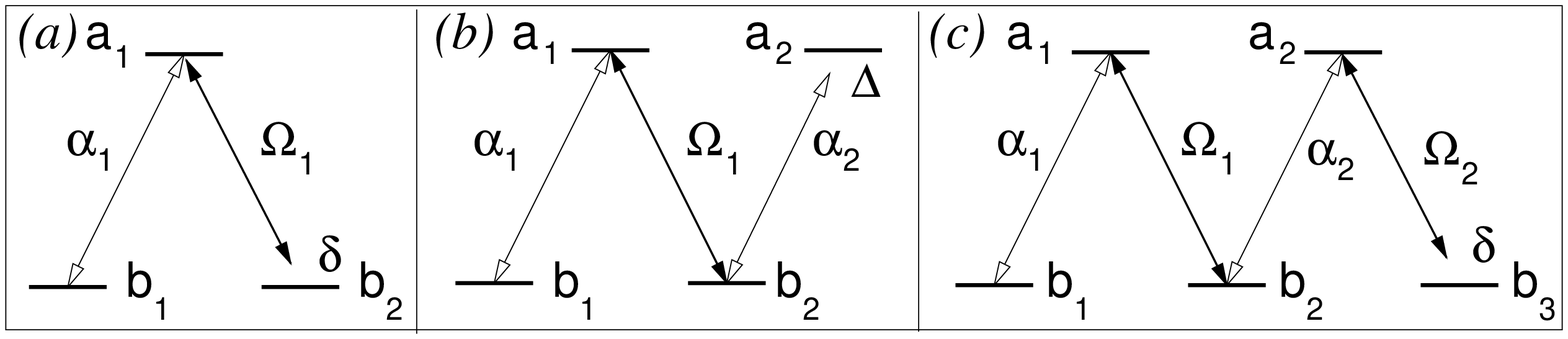, width=8.5cm, angle=0}
 }
\caption{ \label{fig1}
    Energy level schemes for
    (a) $\Lambda$-system;
    (b) $N$-system;
    (c) $M$-system.
}
 \end{figure}

\begin{figure} 
 \center{
\epsfig{file=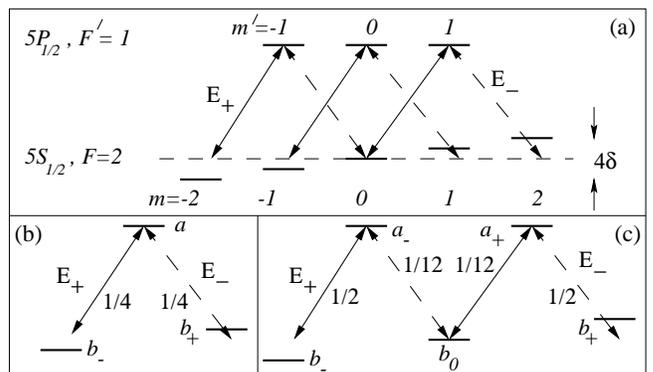, width=8.5cm, angle=0}
 }
\caption{ \label{fig2}
    (a) Energy level scheme for $^{87}$Rb atoms.
    This scheme may be decomposed into a supperposition
    of a $\Lambda$-system (b) and an $M$-system (c).
}
 \end{figure}

\begin{figure} 
 \center{
\epsfig{file=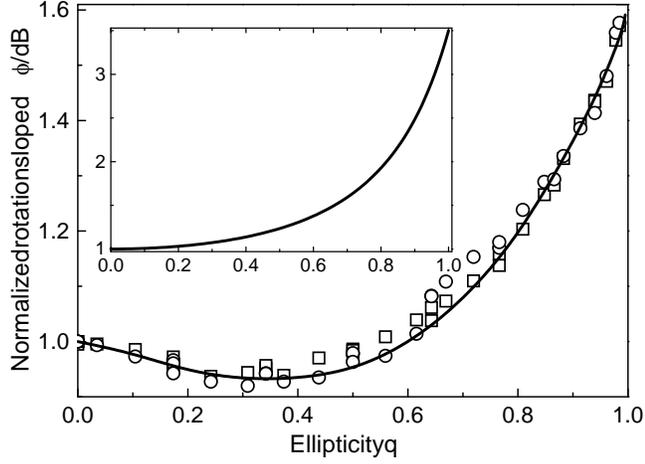, width=6.5cm, angle=90}
 }
\caption{ \label{fig3}
    The slope of nonlinear magneto-optic rotation as a
    function of the ellipticity of the incident light.
    The slope is normalized by the value for linearly
    polarized light (zero ellipticity).  Experimental data
    are shown by squares (1 mW laser power) and (2 mW laser
    power), while the results of the numerical simulations
    for the case of $2$~mW laser power are shown by the
    solid line.  The laser beam has $2$~mm diameter.  Inset:
    the theoretical dependence for naturally broadened Rb
    vapor, from Eq.~(\ref{rot1}).
}
\end{figure}
\end{flushleft}

\end{document}